\input epsf
\catcode`@=11
 \font\tenrm=cmr8 
  \font\sevenrm=cmr7
  \font\fiverm=cmr5
 \font\teni=cmmi8 
  \font\seveni=cmmi7
  \font\fivei=cmmi5
 \font\tensy=cmsy8 
  \font\sevensy=cmsy7
  \font\fivesy=cmsy5
 \font\tenex=cmex8 
 \font\tenbf=cmbx8 
  \font\sevenbf=cmbx7
  \font\fivebf=cmbx5
 \font\tentt=cmtt8 
 \font\tensl=cmsl8 
 \font\tenit=cmti8 
 \skewchar\teni='177 \skewchar\seveni='177 \skewchar\fivei='177
 \skewchar\tensy='60 \skewchar\sevensy='60 \skewchar\fivesy='60
 \textfont0=\tenrm \scriptfont0=\sevenrm \scriptscriptfont0=\fiverm
  \def\rm{\fam\z@\tenrm}
 \textfont1=\teni \scriptfont1=\seveni \scriptscriptfont1=\fivei
  \def\mit{\fam\@ne} \def\oldstyle{\fam\@ne\teni}
 \textfont2=\tensy \scriptfont2=\sevensy \scriptscriptfont2=\fivesy
  \def\cal{\fam\tw@}
 \textfont3=\tenex \scriptfont3=\tenex \scriptscriptfont3=\tenex
 \newfam\itfam \def\it{\fam\itfam\tenit} 
  \textfont\itfam=\tenit
 \newfam\slfam \def\sl{\fam\slfam\tensl} 
  \textfont\slfam=\tensl
 \newfam\bffam \def\bf{\fam\bffam\tenbf} 
  \textfont\bffam=\tenbf \scriptfont\bffam=\sevenbf
  \scriptscriptfont\bffam=\fivebf
 \newfam\ttfam \def\tt{\fam\ttfam\tentt} 
 \normalbaselineskip=12pt
 \setbox\strutbox=\hbox{\vrule height8.5pt depth3.5pt width 0pt}
 \normalbaselineskip\rm
\catcode`@=12

\thinmuskip=2mu
\medmuskip=3mu plus 1mu minus 3mu
\thickmuskip=4mu plus 4mu

\newdimen\fullhsize
\fullhsize=18truecm
\def\fullline{\hbox to\fullhsize}
\def\makeheadline{\vbox to 0pt{\vskip-22.5pt
 \fullline{\vbox to8.5pt{}{\pagefont\the\pageno}\runninghead\hfil}\vss}
 \vskip-5truept\hrule\vskip12truept\nointerlineskip}
\def\makefootline{\baselineskip=12pt\fullline{\the\footline}}
\footline={}

\voffset=-0.3truecm
\vsize=24truecm
\hoffset=-1.1truecm
\hsize=18truecm %
\parindent=10truept
\parskip=0pt
\baselineskip=10truept

%
%
\tolerance=10000
\newdimen\colwidth \newdimen\bigcolheight
\newdimen\pagewidth \newdimen\pageheight
%
%
\colwidth=\hsize
  \advance\colwidth by -.35truein
  \divide\colwidth by 2
\bigcolheight=\vsize
  \advance\bigcolheight by \vsize
\newdimen\savevsizea \savevsizea=\vsize \advance\savevsizea by 24pt
\newdimen\savevsize \savevsize=\vsize
\newdimen\savehsize \savehsize=\hsize
\def\makefootline{\baselineskip=24pt\hbox to \savehsize{\the\footline}}
\font\sevenrm=cmr7 at 7truept
\font\fiverm=cmr5 at 5truept
\font\seveni=cmmi7 at 7truept
\font\fivei=cmmi5 at 5truept
\font\sevensy=cmsy7 at 7truept
\font\fivesy=cmsy5 at 5truept
\def\Footstrut{\hbox{\vrule height6.72pt depth1.92pt width0pt}}
\def\sevenpoint{\def\rm{\fam0\sevenrm}
        \textfont0=\sevenrm \scriptfont0=\fiverm
        \textfont1=\seveni \scriptfont1=\fivei
        \textfont2=\sevensy \scriptfont2=\fivesy
        \textfont3=\tenex \scriptfont3=\tenex
        \normalbaselineskip=8.64truept
        \normalbaselines\rm}
\def\footnote#1{\edef\@sf{\spacefactor\the\spacefactor}#1\@sf
  \insert\footins\bgroup\sevenpoint
  \interlinepenalty=\interfootnotelinepenalty
  \let\par=\endgraf
  \splittopskip=\ht\strutbox
  \splitmaxdepth=\dp\strutbox \floatingpenalty=20000
  \leftskip=0pt \rightskip=\colwidth \advance\rightskip by .35in
        \spaceskip=0pt \xspaceskip=0pt \parindent=1em
  \indent \bgroup\Footstrut #1\aftergroup\Footstrut\egroup
        \let\next}
\pagewidth=\hsize \pageheight=\vsize
\def\onepageout#1{\shipout\vbox{
    \offinterlineskip
    \makeheadline
    \vbox to\savevsizea{#1
        \boxmaxdepth=\maxdepth}
    \makefootline}
    \advancepageno}
  \output{\onepageout{\unvbox255}}
\newbox\partialpage
\def\begindoublecolumns{\begingroup
  \output={\global\setbox\partialpage=\vbox{\unvbox255}}\eject
  \output={\doublecolumnout} \hsize=\colwidth \vsize=\bigcolheight
  \ifvoid\footins\else\advance\vsize by -\ht\footins\fi
  \advance\vsize by -2\ht\partialpage}
\def\enddoublecolumns{\output={\balancecolumns}\eject
  \global\output={\onepageout{\unvbox255}}
  \global\vsize=\savevsize
  \endgroup \pagegoal=\vsize}
\def\doublecolumnout{\dimen0=\pageheight
  \advance\dimen0 by-\ht\partialpage \splittopskip=\topskip
  \ifvoid\footins\setbox0=\vsplit255 to\dimen0\else
   \dimen1=\dimen0
   \advance\dimen1 by-\ht\footins
   \advance\dimen1 by-12pt
   \setbox0=\vbox to \dimen0{\vss\vsplit255 to\dimen1
        \vskip\skip\footins \kern-3pt \unvbox\footins}\fi
  \setbox2=\vsplit255 to\dimen0
  \onepageout\pagesofar
  \global\vsize=\bigcolheight
  \unvbox255 \penalty\outputpenalty}
\def\pagesofar{\unvbox\partialpage
   \wd0=\hsize \wd2=\hsize \hbox to\pagewidth{\box0\hfil\box2}}
\def\Makevrule{\gdef\pagesofar{\unvbox\partialpage
  \wd0=\hsize \wd2=\hsize \hbox to\pagewidth{\box0\hfil\vrule\hfil\box2}}}
\def\balancecolumns{\setbox0=\vbox{\unvbox255} \dimen0=\ht0
  \advance\dimen0 by\topskip \advance\dimen0 by-\baselineskip
  \divide\dimen0 by2 \splittopskip=\topskip
  {\vbadness=10000 \loop \global\setbox3=\copy0
    \global\setbox1=\vsplit3 to\dimen0
    \ifdim\ht3>\dimen0 \global\advance\dimen0 by1truept \repeat}
  \setbox0=\vbox to\dimen0{\unvbox1}
  \setbox2=\vbox to\dimen0{\unvbox3}
  \global\output={\balancingerror}
  \pagesofar}
\newhelp\balerrhelp{Please change the page
                        into one that works.}
\def\balancingerror{\errhelp=\balerrhelp
        \errmessage{Page can't be balanced}
        \onepageout{\unvbox255}}
%

\font\titlefont=cmbx10 at 12truept
\font\sectionfont=cmbx10
\font\subsfont=cmsl8
\font\headfont=cmbxti10
\font\pagefont=cmbx10

\newcount\refcount \refcount=0
\def\cite#1{\global\advance\refcount by 1\relax {\rm [\the\refcount]}}
\def\nocite#1{\global\advance\refcount by 1\relax}
\newcount\sectionnumber \global\sectionnumber=0
\newcount\subsectionnumber \global\subsectionnumber=0
\def\subsection#1{\global\advance\subsectionnumber by 1\relax
 {\smallskip\noindent{\bf \the\sectionnumber.\the\subsectionnumber.}\quad
 {\subsfont #1}}: }
\def\section#1{\global\advance\sectionnumber by 1\relax
  \global\subsectionnumber=0\relax
  {\bigskip\noindent{\sectionfont \the\sectionnumber.\quad #1}}
  \smallskip\par}
\newcount\fignumber \global\fignumber=0
\def\beginfig{\global\advance\fignumber by 1 \relax
  \begingroup\leftskip=10truept\rightskip=10truept}
\def\endfig{\par\endgroup}
\def\figcap#1{{\beginfig
  \noindent{\bf Figure \the\fignumber:}\quad #1\endfig}}

\def\heading#1{{\centerline{\titlefont#1}}}
\def\WhoDidIt#1{{\noindent#1}}
\def\spose#1{\hbox to 0pt{#1\hss}}
\def\simlt{\mathrel{\spose{\lower 3pt\hbox{$\mathchar"218$}}
     \raise 2.0pt\hbox{$\mathchar"13C$}}}
\def\simgt{\mathrel{\spose{\lower 3pt\hbox{$\mathchar"218$}}
     \raise 2.0pt\hbox{$\mathchar"13E$}}}

\def\frac#1#2{{{#1}\over {#2}}}

\def\etal{{\it et al.}}
\def\ie{{\it i.e.}}
\def\eg{{\it e.g.}}

\def\aj#1,{Astrophys.\ J.\ {\bf #1},\ }
\def\ajs#1,{Astrophys.\ J.\ Supp. {\bf #1},\ }
\def\araa#1,{Ann.\ Rev.\ Astron. Astrophys.\ {\bf #1},\ }
\def\aap#1,{Astron.\ \& Astrophys.\ {\bf #1},\ }
\def\mn#1,{Monthly Not.\ Royal Astron.\ Soc.\ {\bf #1},\ }
\def\ppnp#1,{Prog.\ in Part.\ Nucl.\ Phys.\ {\bf #1},\ }
\def\prl#1,{Phys.\ Rev.\ Lett.\ {\bf #1},\ }
\def\prd#1,{Phys.\ Rev.\ {\bf D#1},\ }
\def\prD#1,{Phys.\ Rev.\ {\bf D#1},\ }
\def\sci#1,{Science {\bf #1},\ }
\def\nat#1,{Nature {\bf #1},\ }
\def\pp{\par\hangindent=.125truein \hangafter=1\relax}
\def\ppextra{\par\hangindent=.125truein \hangafter=0\relax}

\hyphenation{astro-ph}
\hyphenation{an-iso-tro-py}
\hyphenation{an-iso-tro-pies}
\hyphenation{quadru-pole}
\hyphenation{temp-era-ture}
\hyphenation{fluc-tua-tions}

%
%
%

\baselineskip=10truept

%
\def\runninghead{{\headfont\quad Cosmic Microwave Background
\hfill Scott \& Smoot}}
\overfullrule=5pt


\null
\heading{COSMIC MICROWAVE BACKGROUND MINI-REVIEW}
\smallskip

\begindoublecolumns
\null
\vskip -1truecm

\WhoDidIt{Revised August 2005 by Douglas Scott (University of British Columbia)
and George F. Smoot (UCB/LBNL) for `The Review of Particle
Physics', S. Eidelman {\it et al.}, Physics Letters, B.{\bf 592}, 1
(2004) and 2005 partial update for the 2006 edition available on the PDG WWW 
pages (URL: http://pdg.lbl.gov/)}

\def\spose#1{\hbox to 0pt{#1\hss}}
\def\simlt{\mathrel{\spose{\lower 3pt\hbox{$\mathchar"218$}}
     \raise 2.0pt\hbox{$\mathchar"13C$}}}
\def\simgt{\mathrel{\spose{\lower 3pt\hbox{$\mathchar"218$}}
     \raise 2.0pt\hbox{$\mathchar"13E$}}}

\section{Introduction}

The energy content in radiation from beyond our Galaxy is dominated by
the Cosmic Microwave Background (CMB), discovered in 1965 \cite{Penzias65}.
The spectrum of the CMB is well described by a blackbody
function with $T=2.725\,$K.
This spectral form is one of the main pillars of the
hot Big Bang model for the early Universe.
The lack of any observed deviations from a blackbody spectrum constrains
physical processes over cosmic history at redshifts
$z\simlt 10^7$ (see previous versions of this mini-review \cite{SmoSco}).
However, at the moment, all viable cosmological models predict a very nearly
Planckian spectrum, and so are not stringently limited.

Another observable quantity inherent in the CMB is the variation in
temperature (or intensity) from one part of the microwave sky to
another \cite{WSS}.
Since the first detection of these anisotropies by the {\sl COBE\/}
satellite \cite{smoot92}, there has been intense activity to map the
sky at increasing levels of sensitivity and angular resolution.  A series
of ground- and balloon-based measurements was joined in 2003 by
the first results from NASA's Wilkinson Microwave Anisotropy Probe
({\sl WMAP\/}) \cite{bennett03}.
These observations have led to a stunning confirmation of
the `Standard Model of Cosmology.'  In combination with other astrophysical
data, the CMB anisotropy measurements place quite precise constraints on
a number of cosmological parameters, and have launched us into an era
of precision cosmology.

\section{Description of CMB Anisotropies}

Observations show that the CMB contains anisotropies at the
$10^{-5}$ level, over a wide range of angular scales.
These anisotropies are usually expressed by using a spherical harmonic
expansion of the CMB sky:
$$
T(\theta,\phi) = \sum_{\ell m} a_{\ell m} Y_{\ell m}(\theta, \phi) .
$$
The vast majority of the cosmological information is contained in the
temperature 2 point function, \ie,~the variance as a function of
separation $\theta$.  Equivalently, the power per unit $\ln\ell$ is
$\ell\sum_m\left|a_{\ell m}\right|^2/4\pi$.

\subsection{The Monopole}

The CMB has a mean temperature of $T_\gamma = 2.725\pm0.001\,$K
($1\sigma$) \cite{mather99},
which can be considered as the monopole component of CMB maps, $a_{00}$.
Since all mapping experiments involve
difference measurements, they are insensitive to this average level.
Monopole measurements can only be made with absolute temperature
devices, such as the FIRAS instrument on the {\sl COBE\/}
satellite [6].  Such measurements of the spectrum are
consistent with a blackbody distribution over more than
three decades in frequency.  A blackbody of the measured temperature
corresponds to $n_{\gamma} = (2\zeta(3)/\pi^2)\, T_\gamma^3 \simeq 411\,
{\rm cm^{-3}}$ and $\rho_{\gamma} = (\pi^2 /15)\, T_\gamma^4 \simeq 4.64
 \times 10^{-34}\, {\rm g}\,{\rm cm^{-3}}
 \simeq 0.260\,{\rm eV}\,{\rm cm^{-3}}$.

\subsection{The Dipole}

The largest anisotropy is in the $\ell=1$ (dipole) first spherical
harmonic, with amplitude $3.346\pm0.017\,$mK [5].  The
dipole is interpreted to be the result of the Doppler shift caused by
the solar system motion relative to the nearly isotropic blackbody
field, as confirmed by measurements of the radial velocities of local
galaxies \cite{courteau00}.  The motion of an observer with velocity
$\beta = v/c$ relative to an isotropic Planckian radiation field of
temperature ${T_0}$ produces a Doppler-shifted temperature pattern
$$
\eqalignno{
T(\theta) &= T_0 (1 - \beta^{2})^{1/2}/(1 - \beta \cos\theta) \cr
&\approx T_0 \, \left(1 + \beta \cos\theta + (\beta^{2}/2) \cos2\theta
+ O(\beta^3)\right ). \cr
}
$$
At every point in the sky one observes a blackbody spectrum, with
temperature $T(\theta)$. The spectrum of the dipole is the differential
of a blackbody spectrum, as confirmed by Ref.~\cite{fixsen94}.

The implied velocity \cite{fixsen96} [5] for the solar
system barycenter is $v = 368\pm 2\,{\rm km}\,{\rm s}^{-1}$, assuming a
value $T_0 = T_\gamma$, towards $(\ell,b) =
(263.85^{\circ}\pm0.10^{\circ}, 48.25^{\circ}\pm0.04^{\circ} $).  Such
a solar system velocity implies a velocity for the Galaxy and the Local
Group of galaxies relative to the CMB. The derived value is $v_{\rm LG}
= 627 \pm 22\,{\rm km}\,{\rm s}^{-1}$ towards $(\ell,b) = (276^{\circ}
\pm  3^{\circ}, 30^{\circ} \pm 3^{\circ} $), where most of the error
comes from uncertainty in the velocity of the solar system relative to
the Local Group.

The dipole is a frame dependent quantity, and one can thus determine
the `absolute rest frame' of the Universe as that in which the CMB
dipole would be zero.  Our velocity relative to the Local Group, as
well as the velocity of the Earth around the Sun, and any velocity of
the receiver relative to the Earth, is normally removed for the
purposes of CMB anisotropy study.

\subsection{Higher-Order Multipoles}

Excess variance in CMB maps at higher multipoles ($\ell \geq 2$) is
interpreted as being the result of perturbations in the density
of the early Universe, manifesting themselves at the epoch of the last
scattering of the CMB photons.  In the hot Big Bang picture, this
happens at a redshift $z\simeq1100$, with little dependence on the
details of the model.  The process by which the hydrogen and helium
nuclei can hold onto their electrons is usually referred to as
recombination \cite{SSS}.  Before this epoch, the CMB photons are
tightly coupled to the baryons, while afterwards they can freely stream
towards us.

Theoretical models generally predict that the $a_{\ell m}$ modes are
Gaussian random fields.  Tests show that this is an extremely good
simplifying approximation \cite{komatsu03}, with only some relatively
weak indications of non-Gaussianity or statistical anisotropy at large scales.
Although non-Gaussianity of various forms is possible in early Universe
models \cite{bartolo04}, the signatures found in existing {\sl WMAP\/} data
are generally considered to be subtle foreground or instrumental artefacts
\cite{deOC} \cite{anomalies}.

With the assumption of Gaussian statistics, and if
there is no preferred axis, then it is the variance of the temperature
field which carries the cosmological information, rather than the
values of the individual $a_{\ell m}$s; in other words the power
spectrum in $\ell$ fully characterizes the anisotropies.  The power at
each $\ell$ is $(2 \ell +1) C_\ell/(4\pi)$, where $C_\ell \equiv
\left\langle{|a_{\ell m}|^2}\right\rangle$, and a statistically
isotropic sky means that all $m$s are equivalent.  We use our
estimators of the $C_\ell$s to constrain their expectation values,
which are the quantities predicted by a theoretical model.  For an
idealized full-sky observation, the variance of each measured $C_\ell$
(\ie,~the variance of the variance) is $[2 /(2 \ell +1 )] C^2_\ell$.  This
sampling uncertainty (known as `cosmic variance') comes about because
each $C_\ell$ is $\chi^2$ distributed with $( 2 \ell +1 )$ degrees of
freedom for our observable volume of the Universe.  For fractional sky
coverage, $f_{\rm sky}$, this variance is increased by $1/f_{\rm sky}$
and the modes become partially correlated.

It is important to understand that theories predict the expectation
value of the power spectrum, whereas our sky is a single realization.
Hence the cosmic variance is an unavoidable source of uncertainty
when constraining models; it dominates the scatter at lower $\ell$s,
while the effects of instrumental noise and resolution dominate at
higher $\ell$s \cite{knox95}.

\subsection{Angular Resolution and Binning}

There is no one-to-one conversion between multipole $\ell$ and
the angle subtended by a particular wavevector projected onto the sky.
However, a single spherical harmonic $Y_{\ell m}$ corresponds to
angular variations of $\theta\sim\pi/\ell$.  CMB maps contain
anisotropy information from the size of the map (or in practice some
fraction of that size) down to the beam-size of the instrument,
$\sigma$.  One can think of the effect of a Gaussian beam as rolling
off the power spectrum with the function ${\rm
e}^{-\ell(\ell+1)\sigma^2}$.

For less than full sky coverage, the $\ell$ modes become correlated.
Hence, experimental results are usually quoted as a series of `band
powers', defined as estimators of $\ell(\ell+1)C_\ell/2\pi$ over
different ranges of $\ell$.  Because of the strong foreground signals
in the Galactic Plane, even `all-sky' surveys, such as {\sl COBE\/} and
{\sl WMAP\/} involve a cut sky.  The amount of binning required to
obtain uncorrelated estimates of power also depends on the map size.

\section{Cosmological Parameters}

The current `Standard Model' of cosmology contains around 10 free
parameters (see `Cosmological
Parameters' mini-review \cite{LahLid}).  The
basic framework is the Friedmann-Robertson-Walker metric (\ie,~a
universe that is approximately homogeneous and isotropic on large
scales), with density perturbations laid down at early times and
evolving into today's structures (see `Big-Bang
Cosmology' mini-review \cite{OliPea}).  These
perturbations can be either `adiabatic' (meaning that there is no
change to the entropy per particle for each species,
\ie,~$\delta\rho/\rho$ for matter is $(3/4)\delta\rho/\rho$ for
radiation) or `isocurvature' (meaning that, for example, matter
perturbations compensate radiation perturbations so that the total
energy density remains unperturbed, \ie,~$\delta\rho$ for matter is
$-\delta\rho$ for radiation).  These different modes give rise to
distinct phases during growth, with those of the adiabatic scenario being
strongly preferred by the data.  Models that generate mainly isocurvature type
perturbations (such as most topological defect scenarios) are no longer
considered to be viable.

Within the adiabatic family of models, there is, in principle, a free
function describing how the comoving curvature perturbations, ${\cal
R}$, vary with scale.  In inflationary models \cite{lidlyth}, the Taylor
series expansion of $\ln{\cal R}(\ln k)$ has terms of steadily decreasing
size.  For the simplest models, there are thus 2 parameters describing
the initial conditions for density perturbations: the amplitude and
slope of the power spectrum, $\left\langle |{\cal
R}|^2\right\rangle\propto k^n$.  This can be explicitly defined, for
example, through:
$$
\Delta_{\cal R}^2\equiv (k^3/2\pi^2)\left\langle |{\cal R}|^2\right\rangle,
$$
and using $A^2\equiv\Delta_{\cal R}^2(k_0)$ with $k_0=0.05\,{\rm
Mpc}^{-1}$, say.  There are many other equally valid definitions of the
amplitude parameter (see also Refs.~[16] and [17]),
and we caution that the relationships between some of them can
be cosmology dependent.  In `slow roll' inflationary models this
normalization is proportional to the combination $V^3/(V^\prime)^2$,
for the inflationary potential $V(\phi)$.  The slope $n$ also involves
$V^{\prime\prime}$, and so the combination of $A$ and $n$ can, in
principle, constrain potentials.

Inflation generates tensor (gravity wave) modes as well as
scalar (density perturbation) modes.
This fact introduces another parameter
measuring the amplitude of a possible tensor component, or equivalently the
ratio of the tensor to scalar contributions.  The tensor amplitude
$A_{\rm T}\propto V$, and thus one expects a larger
gravity wave contribution
in models where inflation happens at higher energies.
The tensor power spectrum also has a slope, often denoted $n_{\rm T}$,
but since this seems likely to be extremely hard to measure, it is sufficient
for now to focus only on the amplitude of the gravity wave component.
It is most common to define the tensor contribution through $r$, the ratio
of tensor to scalar perturbation spectra
at large scales (say $k=0.002\,{\rm Mpc}^{-1}$); however, there are other
definitions in terms of the ratio of contributions to $C_2$, for example.
Different inflationary potentials will lead to different predictions,
e.g.~for $\lambda \phi^4$ inflation with 50 e-folds, $r=0.32$, while other
models can have arbitrarily small values of $r$.
In any case, whatever the
specific definition, and whether they come from inflation or
something else, the `initial conditions' give rise to a minimum of
3 parameters: $A$, $n$, and $r$.

\vskip0.5truecm
\centerline{\epsfxsize=9.0truecm \epsfbox{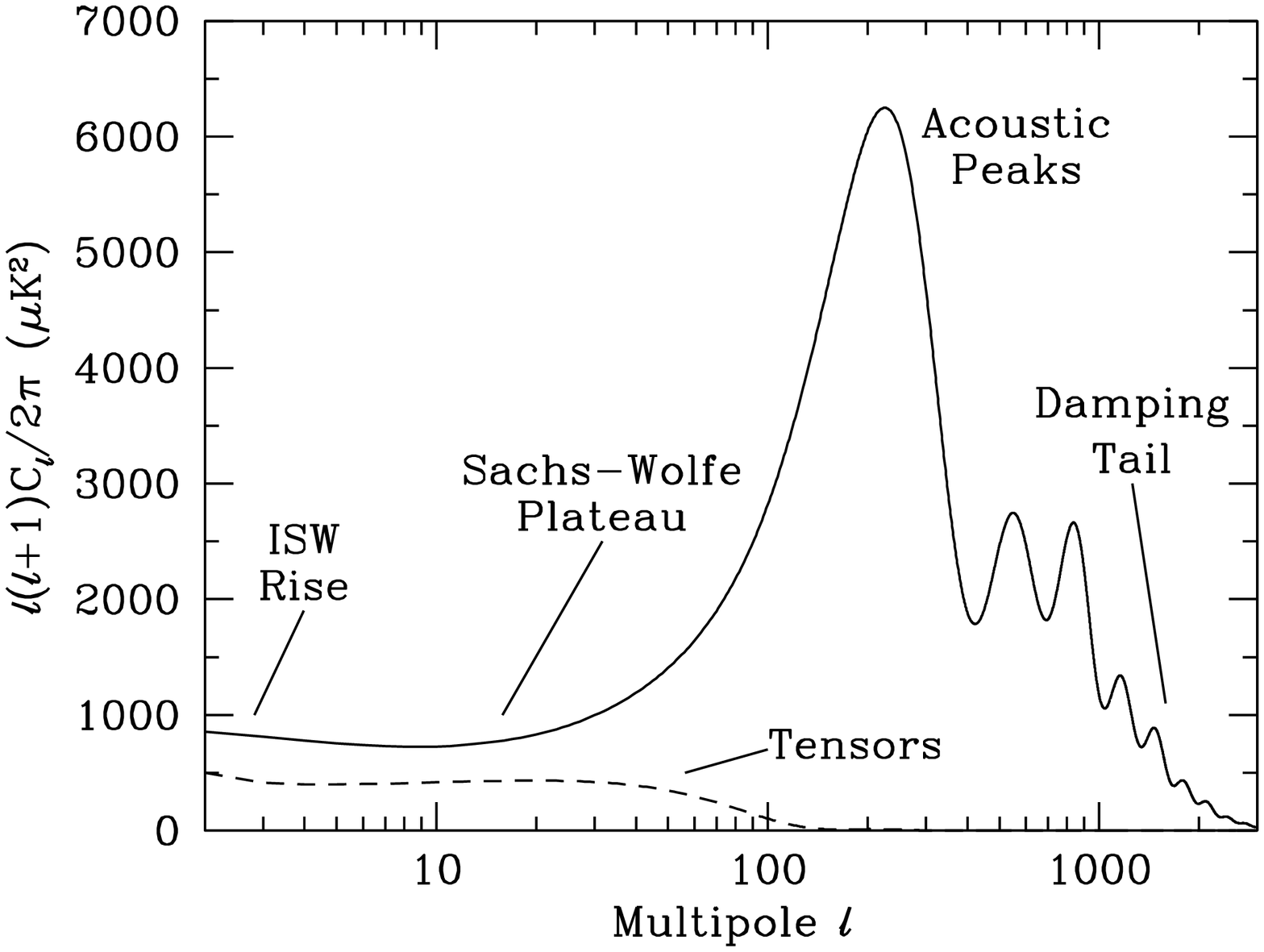}}

\noindent
\figcap{The theoretical CMB anisotropy power spectrum, using
a standard $\Lambda$CDM model from {\tt CMBFAST}.
The $x$-axis is logarithmic
here.  The regions, each covering roughly a decade in $\ell$, are labeled as
in the text: the ISW Rise; Sachs-Wolfe
Plateau; Acoustic Peaks; and Damping Tail.  Also shown is the shape of the
tensor (gravity wave) contribution, with an arbitrary normalization.}
\medskip

The background cosmology requires an expansion parameter (the Hubble
Constant, $H_0$, often represented through
$H_0=100\,h\,{\rm km}\,{\rm s}^{-1}{\rm Mpc}^{-1}$)
and several parameters to describe the matter
and energy content of the Universe.  These are usually given in terms
of the critical density, \ie,~for species `x',
$\Omega_{\rm x}=\rho_{\rm x}/\rho_{\rm crit}$,
where $\rho_{\rm crit}=3H_0^2/8\pi G$.  Since physical densities
$\rho_{\rm x}\propto\Omega_{\rm x}h^2\equiv\omega_{\rm x}$ are what govern the
physics of the CMB anisotropies, it is these $\omega$s that are best
constrained by CMB data.  In particular CMB observations constrain
$\Omega_{\rm B}h^2$ for baryons
and $\Omega_{\rm M}h^2$ for baryons plus Cold Dark Matter.

The contribution of a cosmological constant $\Lambda$ (or other form of Dark
Energy) is usually included through a parameter which quantifies the curvature,
$\Omega_{\rm K}\equiv 1-\Omega_{\rm tot}$, where
$\Omega_{\rm tot}=\Omega_{\rm M}+\Omega_{\Lambda}$.  The radiation content,
while in principle a free parameter, is precisely enough determined by
the measurement of $T_\gamma$, and makes a negligible contribution to
$\Omega_{\rm tot}$ today.

The main effect of astrophysical processes on the $C_\ell$s comes through
reionization.  The Universe became reionized at some redshift $z_{\rm i}$,
long after recombination,
affecting the CMB through the integrated Thomson scattering optical
depth:
$$
\tau = \int_0^{z_{\rm i}} \sigma_{\rm T} n_{\rm e}(z) {dt\over dz} dz,
$$
where $\sigma_{\rm T}$ is the Thomson cross-section, $n_{\rm e}(z)$ is the
number density of free electrons (which depends on astrophysics) and
$dt/dz$ is fixed by the background cosmology.  In principle, $\tau$
can be determined from the small scale matter power spectrum together with the
physics of structure formation and feedback processes.  However, this is
a sufficiently complicated calculation that $\tau$ needs to be considered
as a free parameter.

Thus we have 8 basic cosmological parameters: $A$, $n$, $r$, $h$,
$\Omega_{\rm B}h^2$, $\Omega_{\rm M}h^2$, $\Omega_{\rm tot}$, and $\tau$.
One can add additional parameters
to this list, particularly when using the CMB in combination with other data
sets.  The next most relevant ones might be: $\Omega_\nu h^2$, the massive
neutrino contribution; $w$ ($\equiv p/\rho$), the equation of state
parameter for the Dark
Energy; and $dn/d\ln k$, measuring deviations from a constant spectral index.
To these 11 one could of course add further parameters describing
additional physics, such as details of the reionization process, features
in the initial power spectrum, a sub-dominant contribution of isocurvature
modes, {\it etc}.

As well as these underlying parameters, there are other quantities that can
be obtained from them.  Such derived parameters include
the actual $\Omega$s of the various components (\eg,~$\Omega_{\rm M}$),
the variance of density perturbations at particular scales (\eg,~$\sigma_8$),
the age of the Universe today ($t_0$),  the age of the Universe at
recombination, reionization, {\it etc}.

\section{Physics of Anisotropies}

The cosmological parameters affect the anisotropies through the well
understood physics of the evolution of linear perturbations within a
background FRW cosmology.
There are very effective, fast, and publicly-available software codes 
for computing the CMB anisotropy, polarization, and matter power spectra, 
\eg,~{\tt CMBFAST} \cite{SelZal} and {\tt CAMB} \cite{LewChaLas}.
{\tt CMBFAST} is the most extensively used code;
it has been tested over a wide range of 
cosmological parameters and is considered to be accurate to better than the 1\% 
level \cite{SSWZ}.

A description of the physics underlying the $C_\ell$s can be separated
into 3 main regions, as shown in Fig.~1.

\subsection{The Sachs-Wolfe plateau: $\ell\simlt100$}

The horizon scale (or more precisely, the angle subtended by the Hubble
radius) at last scattering corresponds to $\ell\simeq100$.
Anisotropies at larger scales have not evolved significantly, and hence
directly reflect the `initial conditions.'  The combination of
gravitational redshift and intrinsic temperature fluctuations leads to
$\delta T/T \simeq (1/3) \delta\phi/c^2$, where $\delta\phi$ is the
perturbation to the gravitational potential.  This is usually referred to as
the `Sachs-Wolfe' effect \cite{SacWol}.

Assuming that a nearly scale-invariant spectrum of density perturbations
was laid down at early times (\ie,~$n\simeq1$, meaning equal power per decade
in $k$), then $\ell(\ell+1)C_\ell \simeq {\rm constant}$
at low $\ell$s.  This effect is hard to see unless the multipole axis is
plotted logarithmically (as in Fig.~1, but not Fig.~2).

Time variation of the potentials (\ie,~time-dependent metric perturbations)
leads to an upturn in the $C_\ell$s in the lowest several multipoles;
any deviation from a total equation of state $w=0$ has such an effect.
So the dominance of the Dark Energy at low redshift makes the lowest $\ell$s
rise above the plateau.  This is sometimes called the `integrated Sachs-Wolfe
effect' (or ISW Rise),
since it comes from the line integral of~${\dot \phi}$;
it has been confirmed through correlations between the large-angle
anisotropies and large-scale structure \cite{iswcorr}.
Specific models can also give additional contributions at low~$\ell$
(\eg,~perturbations in the Dark Energy component itself \cite{HuEis}),
but typically these are buried in the cosmic variance.

In principle, the mechanism that produces primordial perturbations could
generate scalar, vector, and tensor modes.  However, the vector (vorticity)
modes decay with the expansion of the Universe.  Tensors also decay when
they enter the horizon, and so they contribute only to angular scales above
about $1^\circ$ (see Fig.~1).
Hence some fraction of the low $\ell$ signal could be
due to a gravity wave contribution, although small amounts of tensors are
essentially
impossible to discriminate from other effects that might raise the level
of the plateau.  However, the tensors {\it can\/} be distinguished
using polarization information (see Sec.~6).

\subsection{The acoustic peaks: $100\simlt\ell\simlt1000$}

On sub-degree scales, the rich structure in the anisotropy
spectrum is the consequence of gravity-driven acoustic oscillations occurring
before the atoms in the Universe became neutral.
Perturbations inside the horizon at last scattering have been able to evolve
causally and produce anisotropy at the last scattering epoch, which reflects
that evolution.
The frozen-in phases of these sound waves imprint a dependence
on the cosmological parameters, which gives CMB anisotropies their great
constraining power.

The underlying physics can be understood as follows.  Before the Universe
became neutral the
proton-electron plasma was tightly coupled to the photons, and these
components behaved as a single `photon-baryon fluid'.
Perturbations in the gravitational potential, dominated by the dark
matter component, were steadily evolving.  They drove oscillations in
the photon-baryon fluid, with photon pressure providing most of the restoring
force and baryons giving some additional inertia.
The perturbations were quite small in amplitude, ${\rm O}(10^{-5})$, and so
evolved linearly. That means each Fourier mode evolved independently and
hence can be described by a driven harmonic oscillator, with frequency
determined by the sound speed in the fluid.  Thus the fluid density
oscillates, which gives oscillations in temperature, together with a
velocity effect which is $\pi/2$ out of phase and has its amplitude
reduced by the sound speed.

After the Universe recombined the baryons and radiation decoupled, and
the radiation could travel freely towards us.  At that point the phases
of the oscillations were frozen-in, and projected on the sky as a
harmonic series of peaks.  The main peak is the mode that went through
1/4 of a period, reaching maximal compression.  The even peaks are
maximal {\it under\/}-densities, which are generally of smaller
amplitude because the rebound has to fight against the baryon inertia.
The troughs, which do not extend to zero power, are partially filled by
the Doppler effect because they are at the velocity maxima.

An additional ingredient comes from geometrical projection.  The scale
associated with the peaks is the sound horizon at last scattering,
which can be straightforwardly calculated as a physical length scale.  This
length is projected onto the sky, leading to an angular scale that
depends on the background cosmology.  Hence the angular position of the
peaks is a sensitive probe of the spatial curvature of the Universe
(\ie,~$\Omega_{\rm tot}$), with the peaks lying at higher $\ell$ in
open universes and lower $\ell$ in closed geometry.

One last effect arises from reionization at redshift $z_{\rm i}$.  A
fraction of photons ($\tau$) will be isotropically scattered at $z<z_{\rm i}$,
partially erasing the anisotropies at angular scales smaller than those
subtended by the Hubble radius at $z_{\rm i}$, which corresponds
typically to $\ell$s above about a few 10s, depending on the specific
reionization model.  The acoustic peaks are therefore reduced by a
factor $e^{-2\tau}$ relative to the plateau.

These peaks were a clear theoretical prediction going back to
about 1970 \cite{PeeYu}.  One can think of them as a snapshot of
stochastic standing waves.  Since the physics governing them is
simple and their structure rich, then one can see how they encode extractable
information about the cosmological parameters.
Their empirical existence started to become
clear around 1994 \cite{scott95}, and the emergence, over the following
decade, of a coherent series of acoustic peaks and troughs is a triumph
of modern cosmology.  This picture has received further confirmation with the
recent detection in the power spectrum of galaxies (at redshifts close to zero)
of the imprint of the acoustic oscillations in the baryon component
\cite{wiggles}.

\subsection{The damping tail: $\ell\simgt1000$}

The recombination process is not instantaneous, giving a thickness to the
last scattering surface.  This leads to a damping of the anisotropies at the
highest $\ell$s, corresponding to scales smaller than that subtended
by this thickness.  One can also think of the photon-baryon fluid as having
imperfect coupling, so that there is diffusion between the two components,
and hence the amplitudes of the oscillations decrease with time.
These effects lead to a damping of
the $C_\ell$s, sometimes called Silk damping \cite{silk}, which cuts off the
anisotropies at multipoles above about 2000.

An extra effect at high $\ell$s comes from gravitational lensing, caused
mainly by non-linear structures at low redshift.  The $C_\ell$s are
convolved with a smoothing function in a calculable way, partially 
flattening the peaks, generating a power-law tail at the highest
multipoles, and complicating the polarization signal \cite{ZalSel}.
This is an example of a `secondary effect', \ie,~the processing of anisotropies
due to relatively nearby structures.  Galaxies and clusters of galaxies
give several such effects, but all are expected to be of low amplitude
and are typically only important for the highest $\ell$s.

\section{Current Anisotropy Data}

There has been a steady improvement in the quality of CMB data that has
led to the development of the present-day cosmological model.
Probably the most robust constraints currently available come from the
combination of the {\sl WMAP\/}
first year data [5] with smaller scale results from
the CBI \cite{cbi} and ACBAR \cite{acbar} experiments.  Newer data from
BOOMERANG \cite{boomerang} and the VSA \cite{vsa} are also useful at smaller
angular scales and give consistent results for parameters.
We plot power spectrum estimates from these five experiments in
Fig.~2.
Other recent experiments, such as ARCHEOPS \cite{archeops},
DASI \cite{dasi} and MAXIMA \cite{maxima}
also give powerful constraints, which are quite consistent with what
we describe below.  There have been some comparisons among
data-sets \cite{WMAXIMA},
which indicate very good agreement, both in maps and in derived power
spectra (up to systematic uncertainties in the overall calibration for
some experiments).  This makes it clear that systematic effects are largely
under control.  However, a fully self-consistent joint analysis
of all the current data sets has not been attempted, one of the reasons being
that it requires a careful treatment of the overlapping sky coverage.

\centerline{\epsfxsize=10.0truecm \epsfbox{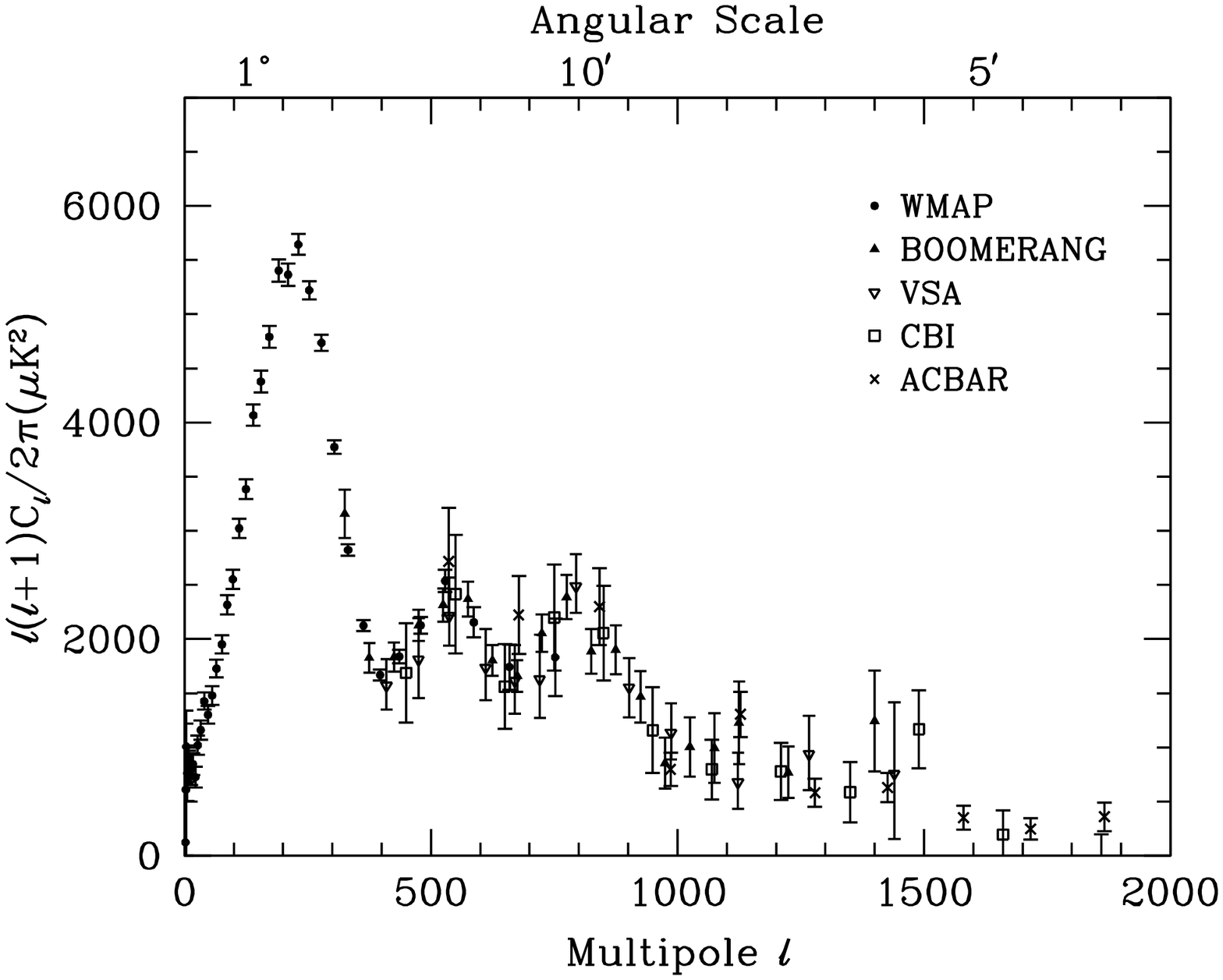}}

\noindent
\figcap{Band-power estimates from the {\sl WMAP}, BOOMERANG, VSA, CBI, and
ACBAR experiments.
We have suppressed some of the low-$\ell$ and high-$\ell$ band-powers which
have large error bars.  Note also that the widths of the $\ell$-bands varies
between experiments.  This plot represent only a selection of available
experimental results, with some other data-sets being of similar quality.
The multipole axis here is linear, so the Sachs-Wolfe plateau is hard to see.
However, the acoustic peaks and damping region are very clearly observed, with
no need for a theoretical curve to guide the eye.}
\smallskip

Fig.~2 shows band-powers from the first year {\sl WMAP\/}
data \cite{hinshaw03}, together with BOOMERANG [32], VSA [33],
CBI [30] and ACBAR [31] data at higher $\ell$.
The points are in very good agreement with a `$\Lambda$CDM' type model,
as described earlier,
with several of the peaks and troughs quite apparent.
For details of how these estimates were arrived at, the strength of
any correlations between band-powers and other information required to
properly interpret them, turn to the original papers.
 
\bigskip
\goodbreak

\section{CMB Polarization}
\nobreak
Since Thomson scattering of an anisotropic radiation field also generates
linear polarization, the CMB is predicted to be polarized at the roughly 5\%
level \cite{hu97}.
Polarization is a spin 2 field on the sky, and the algebra of the modes
in $\ell$-space is strongly analogous to spin-orbit coupling in quantum
mechanics \cite{HuW97b}.
The linear polarization pattern can be decomposed in a number
of ways, with two quantities required for each pixel in a map, often given as
the $Q$ and $U$ Stokes parameters.  However, the most
intuitive and physical decomposition is a geometrical one, splitting
the polarization 
pattern into a part that comes from a divergence (often referred to as
the `E-mode') and a part with a curl (called the `B-mode') \cite{ZSpol}.
More explicitly, the modes are defined in terms of second derivatives of
the polarization amplitude, with the Hessian for the E-modes having principle
axes in the same sense as the polarization, while the B-mode pattern can
be thought of simply as a $45^\circ$ rotation of the E-mode pattern.
Globally one sees that the E-modes have $(-1)^\ell$ parity (like the
spherical harmonics), while the B-modes have $(-1)^{\ell+1}$ parity.

The existence of this linear polarization allows for 6 different cross
power spectra to be determined from data that measure the full temperature
and polarization anisotropy information.
Parity considerations make 2 of these zero, and we are
left with 4 potential observables: $C^{\rm TT}_\ell$, $C^{\rm TE}_\ell$,
$C^{\rm EE}_\ell$, and $C^{\rm BB}_\ell$.
Since scalar perturbations have no handedness,
the B-mode power spectrum can only be generated by vectors or tensors.
Hence, in the context of inflationary models, the determination of a non-zero
B-mode signal is a way to measure the gravity wave contribution (and thus
potentially derive the energy scale of inflation), even if it
is rather weak.  However, one must first eliminate the foreground contributions
and other systematic effects down to very low levels.

The oscillating photon-baryon fluid also results in a series of acoustic
peaks in the polarization $C_\ell$s.  The main `EE' power spectrum has
peaks that are out of phase with those in the `TT' spectrum, because the
polarization anisotropies are sourced by the fluid velocity.  The `TE'
part of the polarization and temperature patterns comes from
correlations between density and velocity perturbations on the last scattering
surface, which can be both positive and negative, and is of larger amplitude
than the EE signal.  There is no polarization
`Sachs-Wolfe' effect, and hence no large-angle plateau.  However, scattering
during a recent period of reionization can create a polarization `bump'
at large angular scales.

The strongest upper limits on polarization
are at the roughly $10\,\mu$K level from the
POLAR \cite{POLAR} experiment at large angular scales and the PIQUE \cite{PIQUE},
COMPASS \cite{COMPASS} and CBI \cite{CBIlimits} experiments at smaller scales.
The first measurement of a polarization signal came in 2002 from the DASI
experiment \cite{dasipol},
which provided a convincing detection, confirming the general paradigm,
but of low enough significance that it lent little
constraint to models.  As well as the E-mode signal, DASI also made a
statistical detection of the TE correlation.

In 2003 the {\sl WMAP\/} experiment demonstrated that is was able to measure
the TE cross-correlation power
spectrum with high precision \cite{kogut03}.  Other recent experimental
results include a weak detection of the EE signal from
CAPMAP \cite{capmap} and more significant detections from CBI \cite{CBIscience},
DASI \cite{leitch05} and BOOMERANG \cite{boomEE}.  The TE signal has also been
detected in several multipole bands by BOOMERANG \cite{boomTE}, and there are
statistical detections by CBI [49] and DASI [50].
Some upper limits on $C_\ell^{\rm BB}$ also exist, but are currently not very
constraining.

The results of {\sl WMAP\/} $C_\ell^{\rm TE}$ are shown in
Fig.~3, along with estimates from the DASI and BOOMERANG experiments.
The measured shape of the cross-correlation power spectrum provides
supporting evidence of the adiabatic nature of the perturbations, as well as
directly constraining the thickness of the last scattering surface.
Since the polarization anisotropies are generated in this scattering
surface, the existence of correlations at angles above about a degree
demonstrate that there were super-Hubble fluctuations at the recombination
epoch.

\centerline{\epsfxsize=10.0truecm \epsfbox{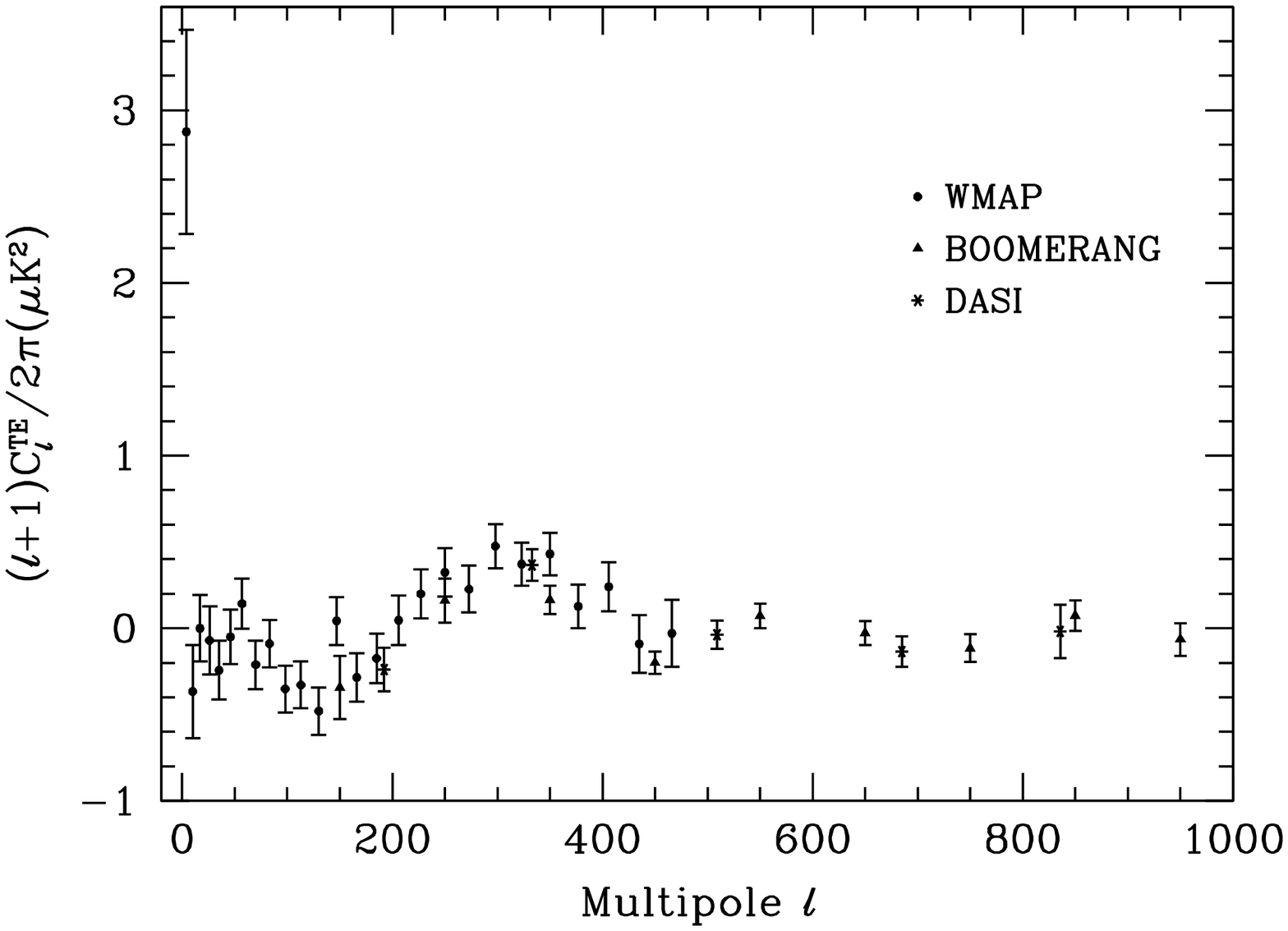}}

\noindent
\figcap
{Cross power spectrum of the temperature anisotropies and
E-mode polarization signal from {\sl WMAP}, together with
estimates from DASI and BOOMERANG which extend to higher $\ell$.
Note that the BOOMERANG bands are wider in $\ell$ than those of {\sl WMAP},
while those of DASI are almost as wide as the features in the power spectrum.
Also note that the $y$-axis here is not
multiplied by the additional $\ell$, which helps to show both the large and
small angular scale features.}
\smallskip

The most intriguing result from the polarization measurements
is at the largest
angular scales ($\ell<10$), where there is an excess signal compared to
that expected from the temperature power spectrum alone.  This is precisely
the signal expected from an early period of reionization, arising from
Doppler shifts during the partial scattering at $z<z_{\rm i}$.
It seems to indicate that the first stars, presumably the source of the
ionizing radiation, formed around $z=20$ (although the uncertainty
is still quite large).

\section{Complications}

There are a number of issues which complicate the interpretation of CMB
anisotropy data, some of which we sketch out below.

\subsection{Foregrounds}

The microwave sky contains
significant emission from our Galaxy and from extra-galactic 
sources \cite{foregrounds}.  Fortunately, the frequency dependence of these 
various sources is in general substantially different from that of the CMB
anisotropy signals.  The combination of Galactic synchrotron, bremsstrahlung
and dust emission reaches a minimum at a wavelength of
roughly $3\,$mm (or about $100\,$GHz).
As one moves to greater angular resolution, the 
minimum moves to slightly higher frequencies, but becomes more sensitive 
to unresolved (point-like) sources.

At frequencies around $100\,$GHz and for portions of the sky away from 
the Galactic Plane the foregrounds are typically 1 to 10\%\ of the CMB 
anisotropies.  By making observations at multiple frequencies, it is relatively
straightforward to separate the various components and determine the CMB
signal to the few per cent level.  For greater sensitivity it is necessary to 
improve the separation techniques by adding spatial information and 
statistical properties of the foregrounds compared to the CMB.

The foregrounds for CMB polarization are expected to follow a 
similar pattern, but are less well studied, and are
intrinsically more complicated.  Whether it is possible to achieve 
sufficient separation to detect B-mode CMB polarization is still an open 
question.  However, for the time being, foreground contamination is not a
major issue for CMB experiments.

\subsection{Secondary Anisotropies}

With increasingly precise measurements of the primary anisotropies, there
is growing theoretical and experimental interest in `secondary anisotropies.'
Effects which happen at $z\ll1000$ become more important as experiments
push to higher angular resolution and sensitivity.

These secondary effects include gravitational lensing, patchy reionization
and the Sunyaev-Zel'dovich (SZ) effect \cite{sunzel80}.  This is
Compton scattering ($\gamma e \rightarrow \gamma' e'$)
of the CMB photons by a hot electron gas, which creates spectral distortions
by transferring energy from the electrons to the photons.
The effect is particularly important for clusters of galaxies, through which
one observes a partially Comptonized spectrum, resulting in a decrement at
radio wavelengths and an increment in the submillimeter.  This can be used to
find and study individual clusters and to obtain estimates of
the Hubble constant.  There is also the potential to constrain
the equation of state of the Dark Energy through counts of detected clusters as
a function of redshift \cite{CHR02}.

\subsection{Higher-order Statistics}

Although most of the CMB anisotropy information is contained in the power
spectra, there will also be weak signals present in higher-order statistics.
These statistics can measure any primordial non-Gaussianity in the
perturbations,
as well as non-linear growth of the fluctuations on small scales and other
secondary effects (plus residual foreground contamination).
Although there are an infinite variety of ways in which the CMB could be
non-Gaussian, there is a generic form to consider for the initial conditions,
where a quadratic contribution to the curvature perturbations is
parameterized through a dimensionless number $f_{\rm NL}$.  This weakly
non-linear component can be constrained through measurements of the
bispectrum or Minkowski functionals for example, and the result from
{\sl WMAP\/} is $-58 < f_{\rm NL} < 134$
(95\% confidence region) [11].

\section{Constraints on Cosmologies}

The clearest outcome of the newer experimental results is that the
standard cosmological paradigm is in good shape.  A large amount of high
precision data on the power spectrum is adequately fit with fewer than 10
free parameters.  The framework is that of Friedmann-Robertson-Walker models,
which have nearly flat geometry, containing Dark Matter and Dark Energy,
and with adiabatic perturbations having close to scale invariant initial
conditions.

Within this framework, bounds can be placed on the values of the
cosmological parameters.  Of course,
much more stringent constraints can be placed on models
which cover a restricted number of parameters, e.g.~assuming
that $\Omega_{\rm tot}=1$, $n=1$ or $r=0$.
More generally, the
constraints depend upon the adopted priors, even if they are implicit,
for example by restricting the parameter freedom or the ranges of
parameters (particularly where likelihoods peak near the boundaries), or
by using different choices of other data in combination with the CMB.
When the data become even more precise,
these considerations will be less important,
but for now we caution that restrictions on model space and choice of
priors need to be kept in mind when adopting specific parameter values and
uncertainties.

There are some combinations of parameters that fit the CMB anisotropies
almost equivalently.  For example, there is a nearly exact geometric
degeneracy,
where any combination of $\Omega_{\rm M}$ and $\Omega_\Lambda$
that gives the same angular diameter distance to last
scattering will give nearly identical $C_\ell$s.  There are also other
near degeneracies among the parameters.  Such degeneracies can be broken
when using the CMB data in combination with other cosmological data sets.
Particularly useful are complementary constraints from galaxy clustering,
the abundance of galaxy clusters, weak gravitational lensing measurements,
Type Ia supernova distances and the distribution of Lyman $\alpha$ forest
clouds.  For an overview of some of these other cosmological constraints,
see Ref.~[16].

The combination of {\sl WMAP\/}, CBI and ACBAR, together with weak priors
(on $h$ and $\Omega_{\rm B}h^2$ for example),
and within the context of a 6 parameter family of
models (which fixes $\Omega_{\rm tot}=1$ and $r=0$),
yields the following results \cite{spergel03}:
$A=2.7(\pm0.3)\times10^{-9}$, $n=0.97\pm0.03$, $h=0.73\pm0.05$,
$\Omega_{\rm B}h^2=0.023\pm0.001$,
$\Omega_{\rm M}h^2=0.13\pm0.01$ and
$\tau=0.17\pm0.07$.
Other combinations of data, e.g.~including BOOMERANG and other newer CMB
measurements, or improved large-scale structure data, lead to consistent
results, sometimes with smaller error bars, and with the precise values
depending on data selection \cite{params}.
Note that for $h$, the CMB data alone provide only a very weak constraint,
unless spatial flatness or some other cosmological data are used.
For $\Omega_{\rm B}h^2$ the precise value depends
sensitively on how much freedom is allowed in the shape of the primordial
power spectrum 
(see `Big-Bang Nucleosynthesis' mini-review \cite{FieSar}).

The best constraint on $\Omega_{\rm tot}$ is $1.02\pm0.02$.  This comes from
including priors from $h$ and supernova data.  However, slightly different, but
consistent results come from using other data combinations.

The 95\% confidence upper limit on $r$ is 0.53 (including some extra
constraint from galaxy clustering).  This limit is stronger if we restrict
ourselves to $n<1$ and weaker if we allow $dn/d\ln k \ne 0$.

There are also constraints on parameters over and above the basic 8 that we
have described.  But for such constraints it is necessary to include
additional data in order to break the degeneracies.  For example the
addition of the Dark Energy equation of state, $w$ adds the partial
degeneracy of being able to fit a ridge in $(w,h)$ space, extending to
low values of both parameters.  This degeneracy is broken when the CMB is
used in combination with independent $H_0$ limits, for example \cite{freedman},
giving $w<-0.5$ at 95\% confidence.  Tighter limits can be placed using
restricted model-spaces and/or additional data.

For the optical depth $\tau$,
the error bar is large enough that apparently quite different results can
come from other combinations of data.
The constraint from the combined {\sl WMAP\/}
$C_\ell^{\rm TT}$ and $C_\ell^{\rm TE}$ data is $\tau=0.17\pm0.04$,
which corresponds (within reasonable models) to a reionization
redshift $9<z_{\rm i}<30$ (95\% CL) [47].
This is a little higher than some theoretical predictions and some indications
{}from studies of absorption in high-$z$ quasar spectra \cite{fan}.
The excitement here is that we have direct information from CMB polarization
which can be combined with other astrophysical measurements to understand
when the first stars formed and brought about the end of the cosmic dark ages.

\section{Particle Physics Constraints}

CMB data are beginning to put limits on parameters which are directly
relevant for particle physics models.  For example there is a limit on
the neutrino contribution $\Omega_{\nu}h^2<0.0076$ (95\% confidence)
{}from a combination of {\sl WMAP\/} and galaxy clustering data from the
2dFGRS project \cite{colless}.  This directly implies a limit on neutrino
mass, $\sum m_{\nu}<0.7\,$eV, assuming the usual number density of fermions
which decoupled when they were relativistic.

A combination of the {\sl WMAP\/} data with other data-sets gives some hint
of a running spectral index, \ie,~$dn/d\ln k \ne 0$ [56].
Although this is still far from resolved \cite{SelMM},
things will certainly improve as new data come in.
A convincing measurement of a non-zero running of the index would be quite
constraining for inflationary models \cite{peiris03}.

One other hint of new physics lies in the fact that the quadrupole
and some of the other low $\ell$ modes seem anomalously low compared with
the best-fit $\Lambda$CDM model [38].
This is what might be expected in a universe which has
a large scale cut-off to the power spectrum, or is
topologically non-trivial.  However, because of cosmic variance, possible
foregrounds, apparent correlations between modes (as mentioned in
Sec.~2.3),
{\it etc}., the significance of this feature is still a matter of
debate [13] \cite{efstathiou03}.

In addition it is also possible to put limits on other pieces of
physics \cite{KamKos},
for example the neutrino chemical potentials, contribution of warm dark matter,
decaying particles,
time variation of the fine-structure constant, or physics beyond
general relativity.  Further particle physics
constraints will follow as the anisotropy measurements increase in
precision.

Careful measurement of the CMB power spectra and non-Gaussianity can in
principle put constraints on physics at the highest energies, including ideas
of string theory, extra dimensions, colliding branes, {\it etc}.  At the moment
any calculation of predictions appears to be far from definitive.  However,
there is a great deal of activity on implications of string theory for
the early Universe, and hence a very real chance that there might be
observational implications for specific scenarios.

\section{Fundamental Lessons}

More important than the precise values of parameters is what we have
learned about the general features which describe our observable Universe.
Beyond the basic hot Big Bang picture, the CMB has taught us that:

\item{$\bullet$}
The Universe recombined at $z\simeq1100$ and started to become ionized again at
$z\simeq10$--30.
\item{$\bullet$}
The geometry of the Universe is close to flat.
\item{$\bullet$}
Both Dark Matter and Dark Energy are required.
\item{$\bullet$}
Gravitational instability is sufficient to grow all of the observed large
structures in the Universe.
\item{$\bullet$}
Topological defects were not important for structure formation.
\item{$\bullet$}
There are `synchronized' super-Hubble modes generated in the early
Universe.
\item{$\bullet$}
The initial perturbations were adiabatic in nature.
\item{$\bullet$}
The perturbations had close to Gaussian (\ie,~maximally random)
initial conditions.

It is very tempting to make an analogy between the status of the cosmological
`Standard Model' and that of particle physics.  In cosmology there are about
10 free parameters, each of which is becoming well determined, and with a great
deal of consistency between different measurements.  However, none of these
parameters can be calculated from a fundamental theory, and so hints of the
bigger picture, `physics beyond the Standard Model' are being searched for
with ever more ambitious experiments.

Despite this analogy,
there are some basic differences.  For one thing, many of the cosmological
parameters change with cosmic epoch, and so the measured
values are simply the ones
determined today, and hence they are not `constants', like particle
masses for example (although they {\it are\/} deterministic, so that if one
knows their values at one epoch, they can be calculated at another).
Moreover, the number of parameters is not as fixed as it is in
the particle physics Standard Model; different researchers will not
necessarily agree on what the free parameters are, and new ones
can be added as the quality of the data improves.  In addition,
parameters like $\tau$, which come from astrophysics,
are in principle calculable from known physical processes, although this
is currently impractical.  On top of all this, other parameters might be
`stochastic' in that they may be fixed only in our observable patch
of the Universe or among certain vacuum states in the `Landscape'
\cite{susskind03}.

In a more general sense the cosmological `Standard
Model' is much further from the underlying `fundamental theory' which
will ultimately provide the
values of the parameters from first principles.  Nevertheless, any
genuinely complete `theory of everything' must include an explanation for
the values of these cosmological parameters as well as the parameters of
the Standard Model of particle physics.

\bigskip
\goodbreak

\section{Future Directions}
\nobreak
With all the observational progress in the CMB and the tying down of
cosmological parameters, what can we anticipate for the future?
Of course there will 
be a steady improvement in the precision and confidence with which we 
can determine the appropriate cosmological model and its parameters. 
We can anticipate that the evolution from one year to four years of {\sl WMAP\/}
data will bring improvements from the increased statistical accuracy and 
{}from the more detailed treatment of calibration and systematic effects.
Ground-based experiments operating at smaller
angular scales will also improve over the
next few years, providing significantly tighter
constraints on the damping tail.  In addition,
the next CMB satellite mission, {\sl Planck\/}, is scheduled for launch
in 2007, and there are even more ambitious projects currently
being discussed.

Despite the increasing improvement in the results, it is also true that
the addition of the latest experiments has not significantly
changed the cosmological model (apart from a suggestion of higher
reionization redshift perhaps).
It is therefore appropriate to ask: what
should we expect to come from {\sl Planck\/} and from other more grandiose
future experiments, including those being discussed as part of the
`Beyond Einstein' initiative?
{\sl Planck\/} certainly has the advantage of high sensitivity and a
full sky survey.  A detailed measurement of the third acoustic peak
provides a good determination of the matter density; this can only be
done by measurements which are accurate relative to the first two peaks
(which themselves constrain the curvature and the baryon density).
A detailed measurement of the damping tail region will also significantly
improve the determination of $n$ and any running of the slope.
{\sl Planck\/} should also be capable of measuring $C_\ell^{\rm EE}$
quite well, providing both a strong 
check on the cosmological Standard Model and extra constraints that will
improve parameter estimation.

A set of cosmological parameters are now known to roughly 10\% accuracy,
and that may seem sufficient for many people.
However, we should certainly demand more of measurements which describe
{\it the entire observable Universe\/}!  Hence a lot of activity in the coming
years will continue to focus on determining those parameters with
increasing precision.  This necessarily includes testing for consistency
among different predictions of the cosmological Standard Model, and searching
for signals which might require additional physics.

A second area of focus will be the smaller scale anisotropies and
`secondary effects.'  There is a great deal of information about structure
formation at $z\ll1000$ encoded in the CMB sky.  This may involve
higher-order statistics as well as spectral signatures.  Such investigations
can also provide constraints on the Dark Energy equation of state, for example.
{\sl Planck}, as well as experiments aimed at the highest $\ell$s, should
be able to make a lot of progress in this arena.

A third direction is increasingly sensitive searches for specific signatures of
physics at the highest energies.
The most promising of these may be the
primordial gravitational wave signals in $C_\ell^{\rm BB}$, which could
be a probe of the $\sim10^{16}$ GeV energy range.
Whether the amplitude of the effect coming from inflation will be detectable
is unclear, but the prize makes the effort worthwhile.

Anisotropies in the CMB have proven to be the premier probe of cosmology 
and the early Universe.  Theoretically the CMB involves well-understood 
physics in the linear regime, and is under very good calculational 
control.  A substantial and improving set of observational data now exists.
Systematics appear to be well understood and not a limiting factor.
And so for the next few years we can expect an increasing amount of
cosmological information to be gleaned from CMB anisotropies, with the
prospect also of some genuine surprises.

\bigskip
\medskip
\goodbreak
\noindent{\bf Acknowledgements}
\smallskip
We would like to thank the numerous colleagues who helped in compiling and
updating this review.
We are also grateful for the diligence of the RPP staff.
This work was partially supported by the Canadian NSERC and the US DOE.

\medskip
\medskip
\noindent{\bf References:}

\parindent=0pt
\parskip=0pt plus 1pt minus 1pt
\def\pp{\par\hangindent=.175truein \hangafter=1\relax}
\def\ppextra{\par\hangindent=.175truein \hangafter=0\relax}

\pp
1.~A.A. Penzias and R. Wilson, \aj142,419 (1965);

\ppextra
R.H. Dicke \etal,
\aj142,414 (1965)

\pp
2.~G.F. Smoot and D. Scott, in `The Review of Particle Physics', D.E.
Groom, \etal, Eur.\ Phys.\ J.\ {\bf C15}, 1 (2000); {\tt astro-ph/9711069};
{\tt astro-ph/9603157}

\pp
3.~M. White, D. Scott, and J. Silk, Ann.\ Rev.\ Astron.\ \& Astrophys.\
{\bf 32},
329 (1994);

\ppextra
W. Hu and S. Dodelson, Ann.\ Rev.\ Astron.\ \& Astrophys.\ {\bf 40}, 171 (2002)

\pp
4.~G.F. Smoot \etal,  \aj396,L1 (1992)

\pp
5.~C.L. Bennett \etal, \ajs148,1 (2003)

\pp
6.~J.C. Mather \etal, \aj512,511 (1999)

\pp
7.~S. Courteau, J.A. Willick, M.A. Strauss, D. Schlegel, and M. Postman,
\aj544,636 (2000)

\pp
8.~D.J. Fixsen \etal, \aj420,445 (1994)

\pp
9.~D.J. Fixsen \etal, \aj473,576 (1996);

\ppextra
A. Kogut \etal, \aj419,1 (1993)

\pp
10.~S. Seager, D.D. Sasselov, and D. Scott, \ajs128,407 (2000)

\pp
11.~E. Komatsu \etal, \ajs148,119 (2003)

\pp
12.~N. Bartolo \etal, Phys.\ Rep.\ {\bf 402}, 103 (2004)

\pp
13.~A. de Oliveira-Costa \etal, \prD69,063516 (2004)

\pp
14.~Eriksen H.K. \etal, \aj605,14 (2004);

\ppextra
P. Vielva \etal, \aj609,22 (2004);

\ppextra
D.L. Larson, B.D. Wandelt, \aj613,L85 (2004);

\ppextra
P. Bielewicz, K.M. Gorski, and A.J. Banday, Monthly Not.\ Royal Astron.\ Soc.\ 
 {\bf 355}, 1283 (2004);

\ppextra
S. Prunet \etal, \prD71,083508 (2005);

\ppextra
Eriksen H.K. \etal, \aj622,58 (2005)

\pp
15.~L. Knox, \prD52,4307 (1995)

\pp
16.~O. Lahav and A.R. Liddle, in `The Review of Particle Physics', S.
Eidelman \etal, Phys. Lett. B{\bf 592}, 1 (2004); {\tt astro-ph/0601168}

\pp
17.~K. Olive and J.A. Peacock, in `The Review of Particle Physics', S.
Eidelman \etal, Phys. Lett. B{\bf 592}, 1 (2004)

\pp
18.~A.R. Liddle and D.H. Lyth, Cosmological Inflation and Large-Scale Structure,
 Cambridge University Press (2000)

\pp
19.~U. Seljak and M. Zaldarriaga, \aj469,437 (1996)

\pp
20.~A. Lewis, A. Challinor, A. Lasenby, \aj538,473 (2000)

\pp
21.~U. Seljak, N. Sugiyama, M. White, M. Zaldarriaga, \prD68,083507 (2003)

\pp
22.~R.K. Sachs and A.M. Wolfe, \aj147,73 (1967)

\pp
23.~M.R. Nolta \etal, \aj608,10 (2004);

\ppextra
S. Boughn and R. Crittenden, Nature, {\bf 427}, 45 (2004);

\ppextra
P. Fosalba, E. Gazta{\~n}aga, and F. Castander, \aj597,L89 (2003);

\ppextra
N. Afshordi, Y.-S. Loh, and M.A. Strauss, \prD69,083524 (2004);

\ppextra
N. Padmanabhan \etal, Phys. Rev. D in press {\tt astro-ph/0410360\rm}

\pp
24.~W. Hu and D.J. Eisenstein, \prD59,083509 (1999)

\ppextra
W. Hu, D.J. Eisenstein, M. Tegmark, M. White, \prD59,023512 (1999)

\pp
25.~P.J.E. Peebles and J.T. Yu, \aj162,815 (1970);

\ppextra
R.A. Sunyaev and Ya.B. Zel'dovich, Astrophysics \& Space Science {\bf7}, 3
 (1970)

\pp
26.~D. Scott, J. Silk, and M. White, \sci268,829 (1995)

\pp
27.~D.J. Eisenstein \etal, Astrophys. J. in press {\tt astro-ph/0501171\rm};

\ppextra
S. Cole \etal, Monthly Not.\ Royal Astron.\ Soc.\ in press,
{\tt astro-ph/0501174\rm}

\pp
28.~J. Silk, \aj151,459 (1968)

\pp
29.~M. Zaldarriaga and U. Seljak, \prD58,023003 (1998)

\pp
30.~T.J. Pearson \etal, \aj591,556 (2003);

\ppextra
A.C.S. Readhead \etal, \aj609,498 (2004)

\pp
31.~M.C. Runyan \etal, \ajs149,265 (2003);

\ppextra
C.L. Kuo \etal, \aj600,32 (2004)

\pp
32.~J.E. Ruhl \etal, \aj599,786 (2003);

\ppextra
S. Masi \etal, Astrophys. J. in press {\tt astro-ph/0507509\rm};

\ppextra
W.C. Jones \etal, Astrophys. J. in press {\tt astro-ph/0507494\rm}

\pp
33.~P.F. Scott \etal, Monthly Not.\ Royal Astron.\ Soc.\ {\bf 341}, 1076 (2003);

\ppextra
Grainge K. \etal, Monthly Not.\ Royal Astron.\ Soc.\ {\bf 341}, L23 (2003);

\ppextra
C. Dickinson \etal, Monthly Not.\ Royal Astron.\ Soc.\ {\bf 353}, 732 (2004)

\pp
34.~A. Benoit \etal, Astronomy \& Astrophysics {\bf399}, L19 (2003);

\ppextra
M. Tristram \etal, Astronomy \& Astrophysics {\bf436}, 785 (2003)

\pp
35.~N.W. Halverson \etal, \aj568, 38 (2002)

\pp
36.~A.T. Lee \etal, \aj561,L1 (2001)

\pp
37.~M.E. Abroe \etal, \aj605,607 (2004);

\ppextra
N. Rajguru \etal, Astrophys. J. in press {\tt astro-ph/0502330\rm}

\pp
38.~G. Hinshaw \etal, \ajs148,135 (2003)

\pp
39.~W. Hu, M. White, New Astron.\ {\bf 2}, 323 (1997)

\pp
40.~W. Hu, M. White, \prD56,596 (1997)

\pp
41.~M. Zaldarriaga and U. Seljak, \prD55,1830 (1997);

\ppextra
M. Kamionkowski, A. Kosowsky, and A. Stebbins, \prD55,7368 (1997)

\pp
42.~B.G. Keating \etal, \aj560,L1 (2001)

\pp
43.~M.M. Hedman \etal, \aj548,L111 (2001)

\pp
44.~P.C. Farese \etal, \aj610,625 (2004)

\pp
45.~J.K. Cartwright \etal, \aj623,11 (2005)

\pp
46.~J. Kovac \etal, Nature, 420, 772 (2002)

\pp
47.~A. Kogut \etal, \ajs148,161 (2003)

\pp
48.~D. Barkats \etal, \aj619,L127 (2005)

\pp
49.~A.C.S. Readhead \etal, Science, {\bf 306}, 836 (2004)

\pp
50.~E.M. Leitch \etal, \aj624,10 (2005)

\pp
51.~T.E. Montroy \etal, Astrophys. J. in press {\tt astro-ph/0507514\rm}

\pp
52.~F. Piacentini \etal Astrophys. J. in press {\tt astro-ph/0507507\rm}

\pp
53.~A. de Oliveira-Costa, M. Tegmark (eds.), Microwave Foregrounds, Astron. Soc.
 of the Pacific, San Francisco (1999)

\pp
54.~R.A. Sunyaev and Ya.B. Zel'dovich, \araa18,537 (1980);

\ppextra
M. Birkinshaw, Phys.\ Rep.\ {\bf 310}, 98 (1999)

\pp
55.~J.E. Carlstrom, G.P. Holder, and E.D. Reese, Ann.\ Rev.\ Astron.\ \&
 Astrophys.\ {\bf 40}, 643 (2002)

\pp
56.~D.N. Spergel \etal, \ajs148,175 (2003)

\pp
57.~M. Tegmark \etal, \prD69,103501 (2004);

\ppextra
C.J. MacTavish \etal, Astrophys. J. in press {\tt astro-ph/0507503\rm};

\ppextra
A.G. S{\'a}nchez \etal, Monthly Not.\ Royal Astron.\ Soc.\ in press,
{\tt astro-ph/0507583\rm}

\pp
58.~B.D. Fields and S. Sarkar, in `The Review of Particle Physics', S.
Eidelman \etal, Phys. Lett. B{\bf 592}, 1 (2004)

\pp
59.~W.L. Freedman \etal, \aj553,47 (2001)

\pp
60.~X. Fan \etal, \aj123,1247 (2002)

\pp
61.~M. Colless \etal, Monthly Not.\ Royal Astron.\ Soc.\ {\bf 328}, 1039 (2001)

\pp
62.~U. Seljak, P. McDonald, and A. Makarov, Monthly Not.\ Royal Astron.\ Soc.\
 {\bf 342}, L79 (2003)

\pp
63.~H.V. Peiris \etal, \ajs148,213 (2003)

\pp
64.~G. Efstathiou, Monthly Notices of the Royal Astronomical Society, {\bf 346},
 L26 (2003)

\pp
65.~M. Kamionkowski and A. Kosowsky, Ann.\ Rev.\ Nucl.\ Part.\ Sci.\ {\bf 49}, 77
 (1999)

\pp
66.~L. Susskind, The Davis Meeting On Cosmic Inflation
 (2003), {\tt hep-th/0302219\rm}

\vfill
\enddoublecolumns

\bye